\documentclass[aps,prd,twocolumn,superscriptaddress]{revtex4-2}
\usepackage{graphicx}
\usepackage{float}
\usepackage{amsmath}
\usepackage{amsfonts}
\usepackage{color}
\usepackage{bm}
\usepackage{bbm}
\usepackage{array}
\usepackage{amssymb}
\usepackage[english]{babel}
\usepackage{soul}
\usepackage{siunitx}           
\usepackage[colorlinks=true,
    linkcolor=red,
    urlcolor=magenta,
    citecolor=blue,
    anchorcolor=blue]{hyperref}
\usepackage[caption=false]{subfig}

\newcommand{\mytoprule}{\noalign{\vskip 1.5pt}\hline\noalign{\vskip 1.5pt}}
\newcommand{\mymidrule}{\noalign{\vskip 1.5pt}\hline\noalign{\vskip 1.5pt}}
\newcommand{\mybottomrule}{\noalign{\vskip 1.5pt}\hline\noalign{\vskip 1.5pt}}

\begin{document}
	\title{Attosecond electron bunch generation by an intense high-order harmonic pulse interacting with a thin target}
	\author{Yang He }
	\affiliation{Xinjiang Key Laboratory of Solid State Physics and Devices,
	School of Physics Science and Technology, Xinjiang University, Urumqi 830017, China}
	\author{Mamat Ali Bake }\email{mabake@xju.edu.cn}
	\affiliation{Xinjiang Key Laboratory of Solid State Physics and Devices,
	School of Physics Science and Technology, Xinjiang University, Urumqi 830017, China}
	\author{Cheng-Qi Zhang}
	\affiliation{
	Key Laboratory of Beam Technology of the Ministry of Education, and School of Physics and Astronomy, Beijing Normal University, Beijing 100875, China}
	\author{Bai-Song Xie}
	\affiliation{
	Key Laboratory of Beam Technology of the Ministry of Education, and School of Physics and Astronomy, Beijing Normal University, Beijing 100875, China}
	\begin{abstract}
	Laser-accelerated electron bunches and the secondary radiation sources they produce exhibit unique temporal resolution for probing ultrafast physical processes due to their ultrashort pulse duration. The inherently short temporal profile of these pulses leads to extremely high peak bunch currents, thereby enabling a wide range of practical applications. In this study, we propose an innovative method for generating such bunch by utilizing high-harmonics generated through laser-plasma interaction as the driving pulse, which subsequently interacts with a thin target to produce an attosecond electron bunch. Using this method, we successfully generated an electron bunch characterized by excellent collimation and an ultra-short duration of approximately 100 attoseconds, representing a substantial reduction in bunch duration. The total bunch charge achieved was 0.38 nC, with an emittance of $4.5 \times 10^{-3} \, \text{mm} \cdot \text{mrad}$ and a divergence angle of approximately $10^\circ$. Moreover, by systematically analyzing the effects of laser intensity and target positioning, we determined an optimized set of simulation parameters. This research establishes a robust foundation for the generation of ultrashort electron bunches and opens new prospects for their application in advanced high-energy and attosecond physics experiments.

	\end{abstract}
	\maketitle
	
	\section{Introduction}
	The rapid advancement of ultrafast laser technology in recent decades has created major opportunities in fundamental physics and engineering \cite{Mourou:2006}. Modern lasers, which generate femtosecond or attosecond pulses, have greatly improved our ability to study and control fast physical processes. The interaction of ultraintense lasers with matter is a promising research area that enables the development of various high-energy, ultrafast radiation sources. These include compact electron sources \cite{Mangles:2004} for imaging  and acceleration \cite{Esarey:2009,Maxson:2017}; high-flux ion sources \cite{Daido:2012} for cancer therapy  and fusion \cite{Linz:2016,Roth:2013,Roth:2001}; positron sources \cite{Ridgers:2014,Chen:2009} for antimatter research and diagnostics \cite{Cassidy:2007,Sarri:2015,Gidley:2006}; and energetic X/$\gamma$-ray sources \cite{Mikhailova:2012,Gizzi:2016} for high-precision atomic and nuclear studies \cite{Carroll:1991}.
	One of the key advantages of these radiation sources is their outstanding temporal and spatial resolution \cite{Brabec:2000,Palffy:2015}. This enables researchers to observe dynamic atomic and subatomic processes that were previously inaccessible through conventional methods. These capabilities have found practical applications in several high-impact scientific domains, including inertial confinement fusion \cite{Tabak:1994,Park:2006}, laboratory astrophysics \cite{Remington:2006}, and material structure probing \cite{Morrison:2014}. Such radiation sources facilitate the investigation of ultrafast electron dynamics, which are essential for the development of next-generation attosecond science.

	The year 2023 marked a milestone in ultrafast science with the award of the Nobel Prize in Physics for the development of attosecond light pulses. These pulses provide unparalleled temporal resolution, thereby enabling the direct observation of electron dynamics in atoms, molecules, and solid-state systems. Particular emphasis has been placed on the generation of attosecond electron bunch sources through the interaction of femtosecond lasers with matter \cite{Ferri:2021,Zhu:2019,Krausz:2009}. Such bunchs, distinguished by their ultrashort duration and high brightness, exhibit considerable potential for applications in ultrafast electron diffraction, time-resolved spectroscopy, and advanced imaging techniques capable of capturing molecular motion in real time  \cite{Sciaini:2011,Piazza:2013}.

	Conventional accelerators can generate electron bunchs at relativistic speeds; however, they are associated with high energy consumption, limited bunch energy, and low bunch density. In laser-plasma interactions, ultra-short electron bunchs can be produced through mechanisms such as $ \bm{J} \times \bm{B} $ \cite{Kruer:1985} and vacuum heating \cite{Brunel:1987}, where the bunch duration is on the order of its wavelength. Alternatively, broad electron bunchs can be compressed in vacuum \cite{Stupakov:2001,Kulagin:2006}; however, this technique typically yields very low charge quantities. Currently, the main approaches for generating high-energy, high-density, and high-charge electron bunchs are Wakefield acceleration \cite{Luttikhof:2010,Gonsalves:2019,Li:2013,Deng:2023} and intense laser-driven interactions with solid thin targets \cite{Borot:2012,Hentschel:2001,Zhou:2021}.

	Significant progress has been achieved in the investigation of laser-driven solid targets for the generation of ultrashort electron bunchs, particularly with respect to targets of varying densities and morphologies, as well as different laser modes, including conventional Gaussian bunchs and vortex bunchs \cite{Hu:2018attosecond,Hu:2018dense,Liseykina:2010,Zhang:2022,Ju:2024,Sun:2024}. In 2022, Zhang et al. \cite{Zhang:2022} demonstrated that the direct interaction of a conventional laser pulse with a nanometer-scale thin target can produce attosecond electron sheets, which serve as electron bunch sources for subsequent nonlinear Compton scattering. However, the resulting electron bunch exhibited a relatively long duration of approximately 800 attoseconds. When colliding with a counter-propagating laser pulse, this bunch generated a $\gamma$-ray source with a similar duration (800 as). The prolonged electron pulse duration constrains its practical applicability, while the experimental challenges associated with aligning dual laser bunchs further complicate implementation.

	Vortex bunchs, due to their intrinsic longitudinal orbital angular momentum, have expanded the scope of research in laser-driven particle acceleration and radiation source generation. In 2024, Ju et al. \cite{Ju:2024} employed a Laguerre-Gaussian bunch to irradiate a thin-film target. Owing to the strong longitudinal electric field component, they obtained an isolated electron bunch with a duration of approximately 400 attoseconds, which subsequently generated ultra-bright gamma rays through synchrotron radiation. In the same year, Sun et al. \cite{Sun:2024} utilized a spatiotemporal optical vortex (STOV) bunch to drive a nanowire target, producing a relativistic isolated attosecond electron sheet with a duration of 670 as and a charge of 65 pC.

	Currently, the durations of ultrashort electron bunchs in such studies typically remain in the range of several hundred attoseconds, primarily due to the micron-scale wavelengths of the driving lasers, which significantly limit their applicability in fields such as electron imaging. Moreover, as a source for nonlinear Compton scattering, the quality of the electron bunch critically influences the characteristics of the resulting gamma-ray photons. Researchers are actively pursuing strategies to overcome this limitation, with the aim of generating ultrashort electron bunchs with higher energies and greater charges. Such advancements would carry significant implications for the development of ultrafast electronics and high-energy-density physics.

	In this paper we proposes a novel approach that utilizes laser-plasma reflection to generate high-order harmonics, which subsequently interact with thin targets to produce ultrashort electron bunchs. The relativistic oscillating mirror mechanism converts the original low-frequency monochromatic laser into broadband multi-frequency radiation. Through Fourier transform analysis comparing the electric field spectra before and after reflection, significant spectral broadening of the reflected pulse is demonstrated. This frequency up-conversion serves as the fundamental mechanism enabling the generation of electron bunchs with substantially reduced durations. The resulting ultrashort electron bunch exhibits a cutoff energy of $35\,\text{MeV}$, a duration of $100$ as, an emittance of $4.5 \times 10^{-3} \, \text{mm} \cdot \text{mrad}$, and a charge of $0.38\,\text{nC}$. Characterized by its extremely short bunch duration, high charge, and excellent collimation, this electron bunch not only provides a foundation for realizing high-quality attosecond electron bunchs and gamma radiation sources, but also opens new avenues for the study of laser-driven ultrashort particle acceleration and its applications. Furthermore, it offers robust technological support for cutting-edge research in fields such as ultrafast science, particle physics, medical imaging, and materials science.

\section{model setup and simulation parameters}\label{SecII}
\begin{figure}[ht]\suppressfloats
	\includegraphics[scale=0.85]{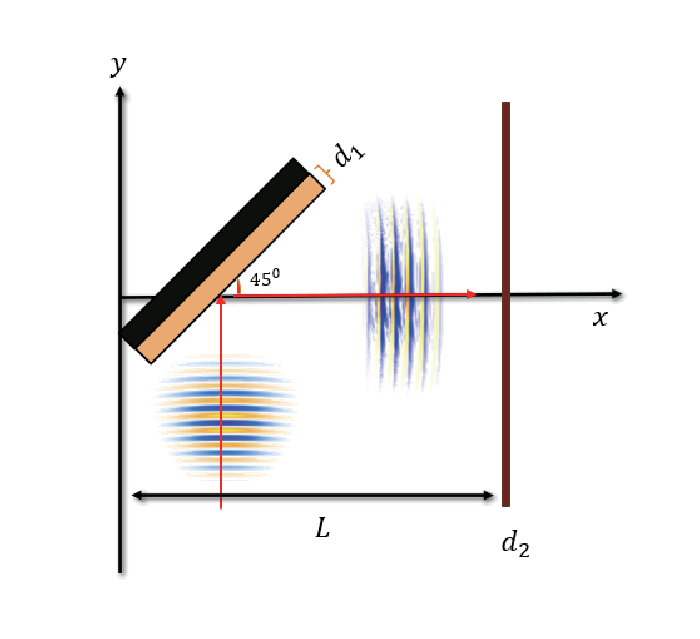}
	\caption{The setup of the attosecond electron bunch generation scheme. A linearly polarized laser bunch enters the simulation region from the position $(5, -10)~\si{\micro\meter}$ and propagates along the positive $y$-axis. At the point $(5, 0)~\si{\micro\meter}$
	, it is reflected from the preplasma target $d_1$, generating a high-order harmonic (HH) pulse. Subsequently, the HH pulse interacts with a uniform thin target $d_2$ located at $\SI{15}{\micro\meter}$, resulting in the production of an attosecond electron bunch.}
	\label{fig1}
\end{figure}

To investigate high-order harmonic generation (HHG) and attosecond electron bunch generation, we conduct two-dimensional particle-in-cell (2D PIC) simulations using the open-source code EPOCH \cite{Arber:2015}, which accurately models laser-plasma interactions.
As shown in Figure.~\ref{fig1}, a linearly polarized (LP) laser pulse enters the simulation domain from the negative $y$-region and propagates along the $y$-direction, with its optical axis offset by \SI{5}{\micro\meter} from the $y$-axis. The pulse subsequently interacts with a pre-plasma target inclined at $45^{\circ}$. Following reflection, high-order harmonics are generated and propagate along the positive $x$-direction, where they interact with a thin target to produce an attosecond electron bunch, as shown in Figure \ref{fig1}.
The incident laser is a linearly polarized (LP) laser with a Gaussian intensity profile given by
$$a = a_0 \exp\left[-\frac{(t-\tau)^2}{\sigma_t^2}\right] \exp\left[-\frac{(x-x_0)^2}{\sigma_x^2}\right],$$
where $a_0 = eE_0/(m_e c \omega_0)=30$ denotes the dimensionless laser amplitude, which corresponds to a laser intensity of
$ I_0 \approx \SI{1.2e21}{\watt\per\centi\meter\squared} $
The laser wavelength is denoted by $\lambda_0 = c \mathrm{T_0} = \SI{1}{\micro\meter}$
the focal spot size is represented by $ w = 3\lambda_0 $,
and $ \omega_0 = 2\pi / \mathrm{T_0} $ refers to the angular frequency.
Here, $ -e $ and $ m_e $ denote the electron charge and rest mass, respectively;
$ E_0 $ is the electric field amplitude,
$ c $ is the speed of light in vacuum,
$ \tau = \SI{15}{\femto\second} $ represents the laser pulse duration,
and $x_0=\SI{5}{\micro\meter}$.
The parameters $ \sigma_t = \tau / \sqrt{2\ln(2)} $ and $ \sigma_x = w / \sqrt{2\ln(2)} $ describe the temporal and spatial standard deviations of the Gaussian envelope, respectively.

The high-order harmonic generation target configuration features a \SI{15}{\micro\meter} longitudinal structure composed of two distinct regions: a pre-plasma region with \SI{0.3}{\micro\meter} thickness exhibiting an exponential density profile defined as $100n_c \exp[(l-d_1)/l_0]$, where $d_1 = \SI{0.3}{\micro\meter}$ corresponds to the maximum density $100n_c$, $l$ denotes the distance from the midpoint to the right edge, and $l_0=$ \SI{0.1}{\micro\meter} represents the density scale length; and a uniform plasma region with \SI{1}{\micro\meter} thickness and a constant density of $100n_c$. The harmonic pulse interaction specifically occurs at $L = \SI{15}{\micro\meter}$ within a $d_2 = \SI{0.1}{\micro\meter} $ thick layer with an electron density of $40n_c$, where the critical plasma density is defined as $n_c = m_e \omega_0^2 / (4\pi e^2) = \SI{1.1e21}{\per\cubic\centi\meter}$.
The simulation box spans $30\lambda_0 \times 20\lambda_0$ and is discretized into $3000 \times 2000$ cells along the $x$ and $y$ directions, respectively, with each cell containing 20 macro-particles. Absorbing boundary conditions are applied to both the laser field and particles in the transverse direction to minimize numerical reflections.

\section{Numerical Results}\label{Results}
	\subsection{Generation of high-order harmonic pulse}
		\begin{figure}[ht]\suppressfloats
			\includegraphics[scale=0.5]{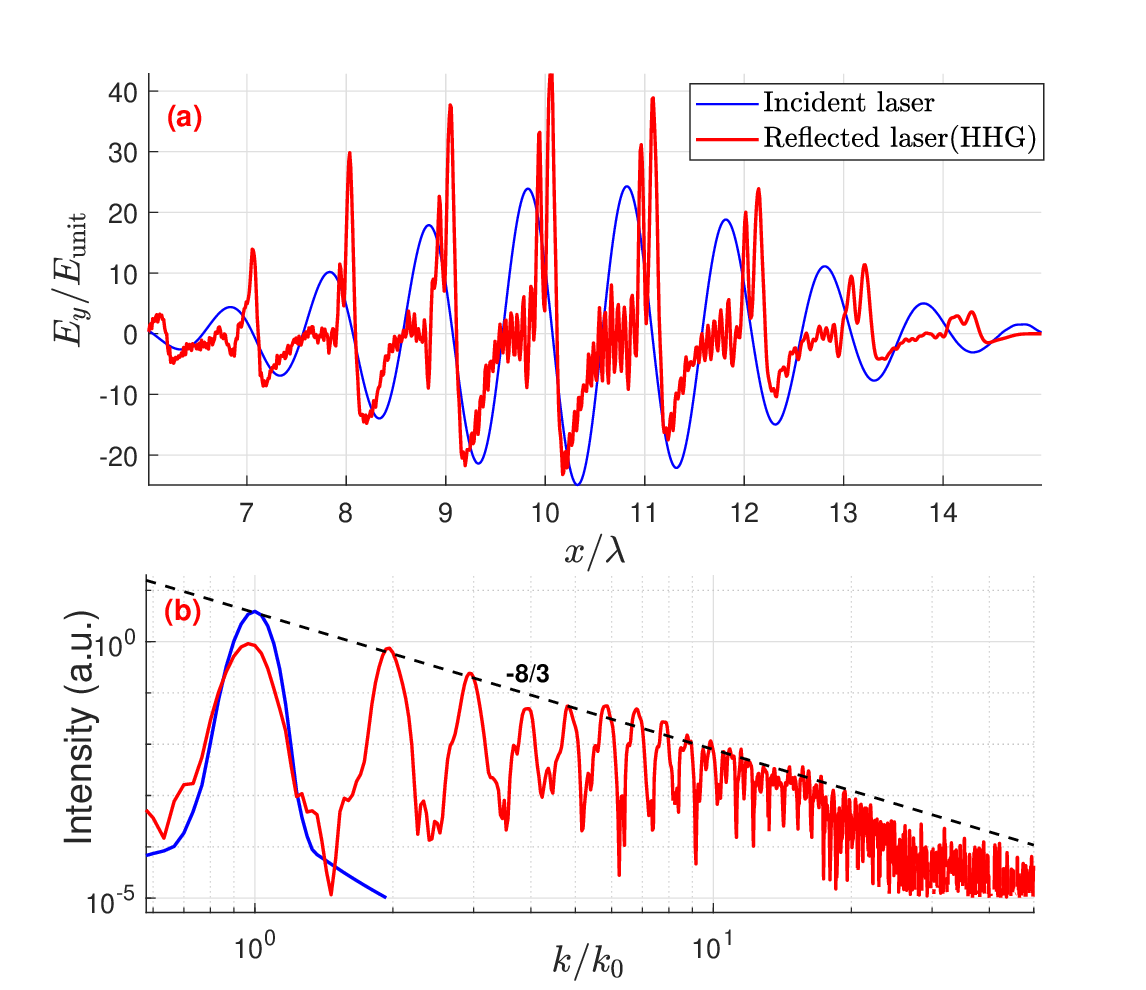}				\caption{(a) The electric field on the optical axis before and after reflection. The blue curve represents the incident electric field before reflection, while the red curve illustrates the reflected electric field containing higher harmonics.
(b) Fourier spectra of the on-axis electric field before and after reflection, with the horizontal axis normalized to the laser frequency. The blue curve corresponds to the incident laser spectrum, the red curve represents the reflected high-order harmonic spectrum, and the black dashed line serves as a calibration reference with a slope of $ -8/3 $.}
			\label{fig2}
		\end{figure}
		Currently, there are two primary approaches for generating higher-order harmonics via laser-matter interaction. The first approach employs ultra-short optical pulses that interact with gaseous media to produce gas-based high-order harmonics, whereas the second approach makes use of ultra-short and ultra-intense lasers that interact with solid targets to generate solid-based high-order harmonics.

 		There exist three primary mechanisms through which lasers and high-density plasmas generate high-order harmonics: coherent wake emission (CWE) \cite{quere:2006}, relativistically oscillating mirror (ROM) \cite{Bulanov:1994}, and coherent synchrotron emission (CSE) \cite{Pukhov:2010}. The dominance of each mechanism depends on the specific laser intensity and plasma conditions. In the context of our setup, the normalized laser intensity exceeds the relativistic threshold. The characteristic scaling behavior of high-harmonic generation intensity clearly indicates that the relativistic oscillating mirror effect is the predominant mechanism.

		The incident laser is entirely reflected at $t=20\mathrm{T_0}$, and according to the law of reflection, the optical axis of the reflected higher-order harmonics lies along the $x$-axis. The electric field on the optical axis is extracted before and after reflection, and the results are presented in Figure.~\ref{fig2}. The electric field distribution curve in Figure.~\ref{fig2}(a). clearly illustrates the significant nonlinear transformation of the incident laser (blue curve) after reflection from the plasma (red curve): the initially smooth, periodic waveform evolves into a complex structure with pronounced subperiodic features, and its peak electric field intensity increases by a factor of $2.3$ compared to the incident field. This transformation results in a steep field gradient at a specific spatial range ($x/\lambda \approx$9--12). This dramatic change in the spatiotemporal structure directly corresponds to the spectral broadening observed in Figure.~\ref{fig2}(b). Fourier analysis of the electric field at $t=20\mathrm{T_0}$, following complete laser reflection, reveals that the spectrum of the reflected laser extends to frequencies several times higher than the fundamental frequency. Notably, in the high-frequency regime ($k/k_0 > 3$), the spectrum exhibits a distinct power-law decay consistent with the $-8/3$ scaling, which aligns well with the theoretical predictions of the relativistic oscillating mirror mechanism \cite{Gordienko:2005}. A comparative analysis of both panels demonstrates that the subwavelength-scale field enhancement structures formed during the reflection process constitute the primary physical mechanism responsible for high-harmonic generation. Moreover, the characteristic power-law spectral decay provides essential diagnostic evidence for studying nonlinear plasma dynamics.

\subsection{Production of an attosecond enectron bunch}
		\begin{figure}[ht]\suppressfloats
			\includegraphics[scale=0.65]{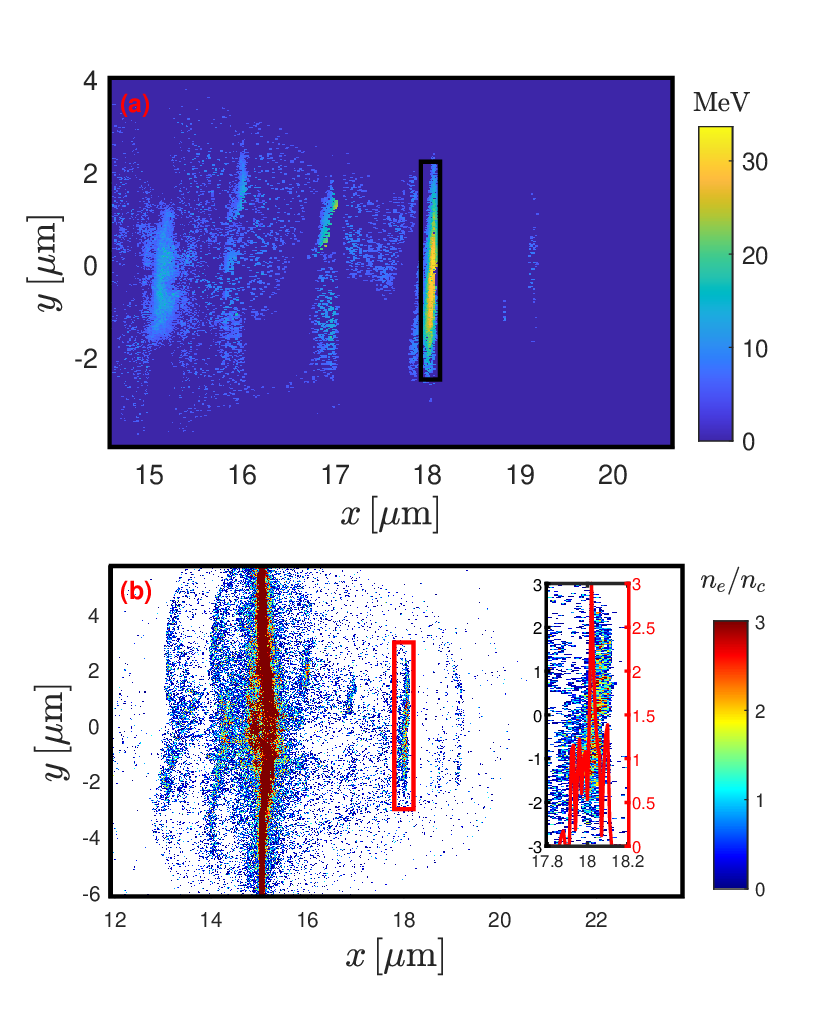}
			\caption{(a) Spatially averaged electron energy at $27\mathrm{T_0}$, with the color bar indicating energy values in \SI{}{\mega\electronvolt}. (b) Spatial distribution of the electron number density at the same time point, with the color bar representing the normalized electron bunch density ($n_{e}/n_{c}$). The red rectangle highlights the region defined by $17.8~\si{\micro\meter} \leq x \leq 18.2~\si{\micro\meter}$ and  $-3~\si{\micro\meter} \leq y \leq 3~\si{\micro\meter}$. The redcurve in the inset displays the electron density profile along $y = 0$ within this selected region.}
			\label{fig3}
		\end{figure}
		Upon completion of the reflection process at $t = 20\mathrm{T_0}$, the reflected high-order harmonic pulse proceeds to interact with a thin plasma target consisting of electrons and aluminum ions positioned at $x=\SI{15}{\micro\meter}$. Owing to their significantly greater mass, the ions remain virtually immobile throughout the interaction, whereas the electrons are accelerated forward by the ponderomotive force exerted by the reflected high-harmonic pulse, leading to the formation of a charge-separation field. In order to expel the electrons from the thin target and generate a relativistic electron sheet, the electric field strength, electron density, and target thickness must fulfill the following condition: $ a_0 \gg \pi \left( \frac{n_e}{n_c} \right) \left( \frac{d_2}{\lambda_0} \right) $, wherein the relevant parameters have been assigned specific numerical values in the model section (see Sec. \ref{SecII}).

		The high-order harmonics pulse exhibit more intense electric field variations and stronger field gradients compared to the original electric field [see Figure~\ref{fig2}(a)], enabling electron acceleration to near-light speeds through the ponderomotive force. Analysis of the planar electron number density distribution indicates the generation of isolated electron bunchs during the interaction process. Time-averaged energy measurements show that high-energy electron bunchs produced through interaction with the thin target propagate consistently along the $x$-direction, maintaining excellent collimation until dispersion occurs as a result of harmonic intensity attenuation at later stages. At the specific simulation time of $t = 27\mathrm{T_0}$, Figure.~\ref{fig3}(a), shows a distinct \SI{35}{\mega\electronvolt} electron bunch within the simulation domain, while Figure.~\ref{fig3}(b). displays the corresponding spatial electron density distribution concentrated around $1~n_{c}$, with rectangular regions ($17.8~\si{\micro\meter} \leq x \leq 18.2~\si{\micro\meter}$, $-3~\si{\micro\meter} \leq y \leq 3~\si{\micro\meter}$) marking the analyzed electron populations.
		\begin{figure}[ht]\suppressfloats
			\includegraphics[scale=0.52]{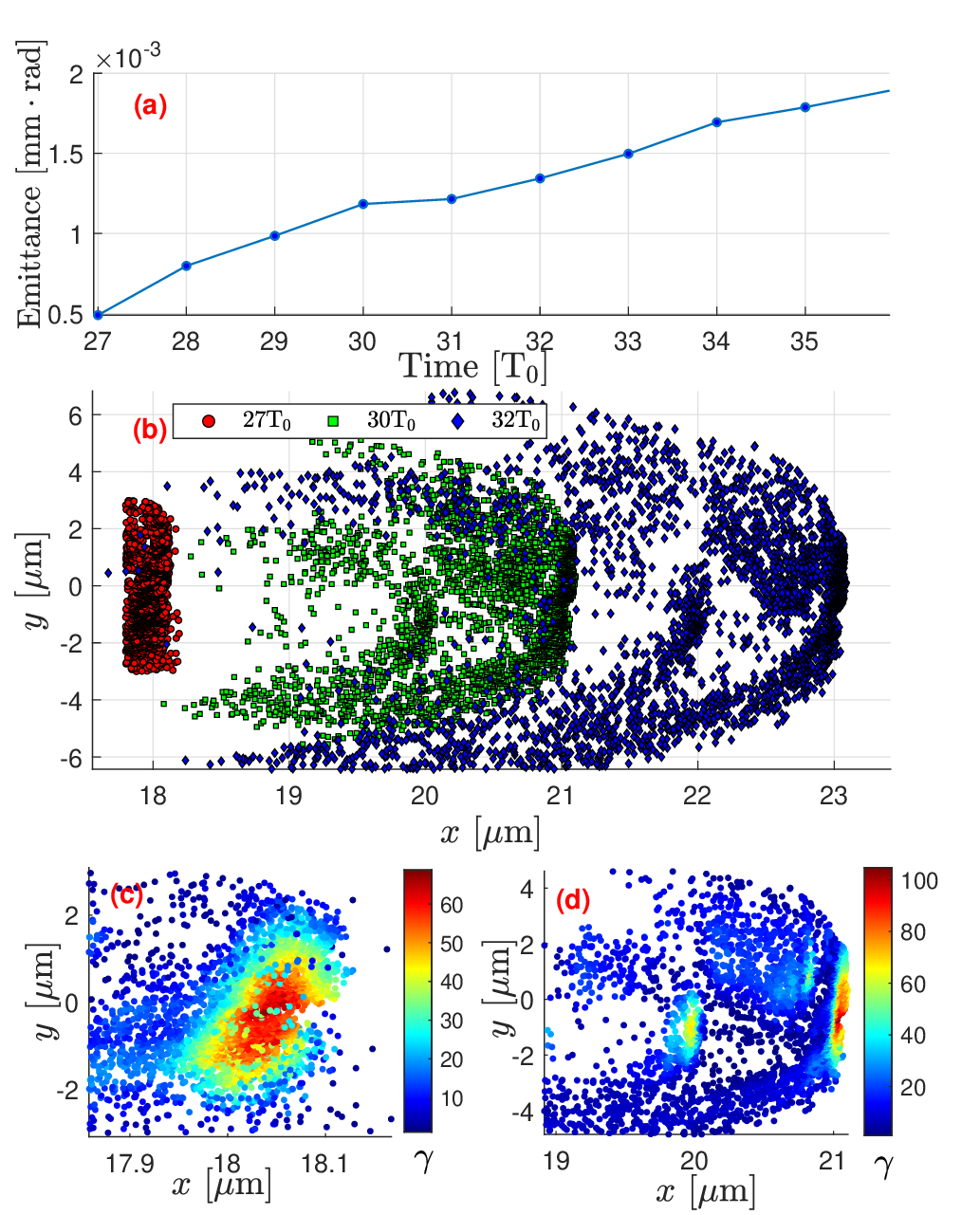}
			\caption{(a) Temporal evolution of electron emittance at $27T_0$ for particles confined within the spatial domain $17.8~\si{\micro\meter} \leq x \leq 18.2~\si{\micro\meter}$ and $-3~\si{\micro\meter} \leq y \leq 3~\si{\micro\meter}$. The plot illustrates the variation of emittance over time for this electron population.
			(b) Scatter plot depicting the spatial distribution of electrons within this region across different time points, where red circles, green squares, and blue diamonds represent macro-electrons at $t=27\mathrm{T_0}$,$t = 30\mathrm{T_0}$ and  $t = 32\mathrm{T_0}$, respectively.
			(c) Spatial distribution of the Lorentz factor ($\gamma$) of electrons at $t=27\mathrm{T_0}$ and $t=30\mathrm{T_0}$, with the color bar indicating the magnitude of $\gamma$.}
			\label{fig4}
		\end{figure}

		\begin{figure*}[ht]\suppressfloats
			\includegraphics[scale=0.8]{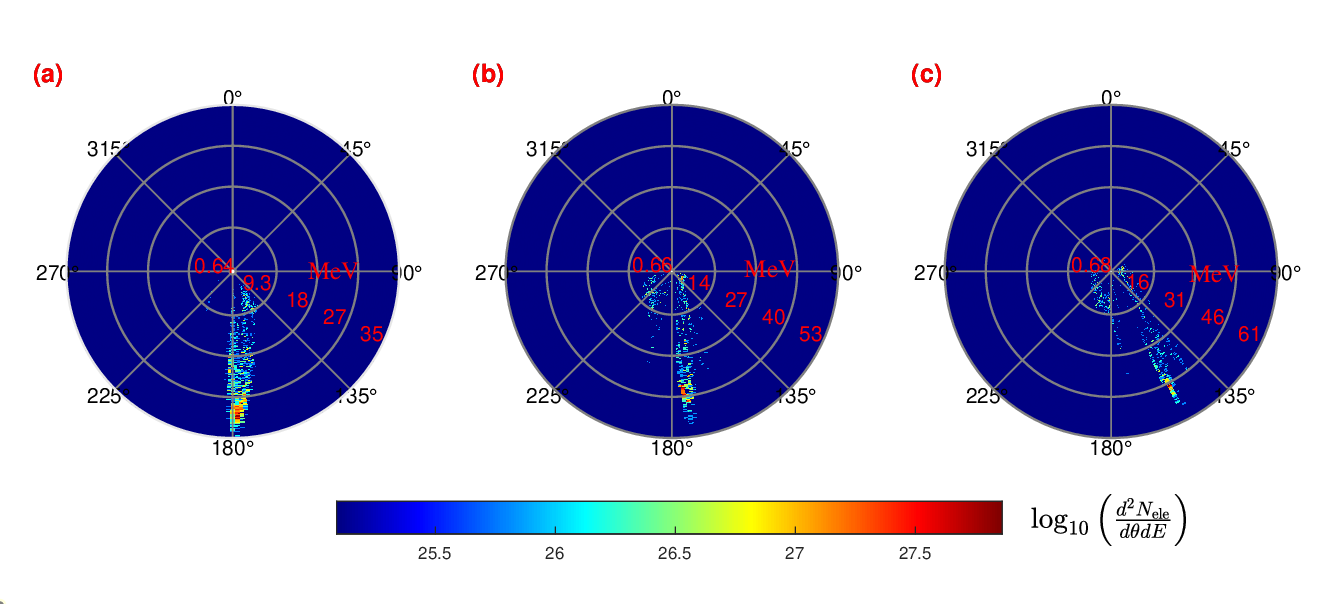}
				\caption{(a), (b), and (c) display the angular energy distributions of electrons at $t=27\mathrm{T_0}$,$t = 30\mathrm{T_0}$ and  $t = 32\mathrm{T_0}$, respectively. The $180^\circ$ direction corresponds to the laser propagation direction ($x$-axis), while the radial direction indicates the electron energy. The color bar represents the logarithmic value of the particle number per unit angle per unit energy, i.e., $\log_{10}\left(\frac{d^{2}N_{\text{ele}}}{d\theta \, dE}\right)$.}
			\label{fig5}
		\end{figure*}

		 To accurately represent the physical charge distribution, the $z$-dimension was scaled by multiplying it with the laser focal spot radius $w$. Accordingly, electrons within the range $y \in [-3, 3]~\si{\micro\meter}$ were selected for further analysis. Statistical analysis of the electrons within the defined region yielded a total electron count of $N = 2 \times 10^9$, corresponding to a total charge of $Q = \SI{0.38}{\nano\coulomb}$. Compared to picoampere-level attosecond electron bunchs reported in previous studies \cite{Sun:2024}, the electron bunch generated in this work exhibits a substantial increase in charge magnitude. In the inset of Figure.~\ref{fig3}(b), electrons within the rectangular region were selected, and their number density distribution along the transverse spatial coordinate at $y=0$ was plotted. The resulting plot displays a distinct peak with an electron number density of approximately $3n_c$. The transverse spatial extent of approximately \SI{0.03}{\micro\meter} corresponds to an electron bunch duration of $100$~fs, which constitutes a key outcome of this investigation.
		 The conventional Gaussian laser bunch undergoes spectral broadening through the relativistic oscillating mirror (ROM) mechanism, where the high-frequency components play a crucial role in generating ultrashort electron bunches. The detailed physical rationale for this phenomenon will be systematically analyzed in subsequent chapters (see Sec.\ref{SecIV--cause} ).

		While the temporal duration of electron bunches remains our primary focus, the collimation characteristics of ultrashort electron bunchs also impose stringent requirements for practical applications. Conventionally described by the angular deviation from the propagation axis ($x$-axis), we employ the emittance parameter to achieve more precise quantification of bunch collimation. The vertical emittance $\varepsilon_y = \sqrt{\langle y^{2} \rangle \langle \theta^{2} \rangle - \langle y \theta \rangle^{2}}$ quantifies the intrinsic phase-space volume of the electron bunch, where $y$ represents the transverse coordinate, $\theta$ denotes the divergence angle relative to the $x$-axis, with $\langle y^{2} \rangle = \frac{1}{N_{\text{ele}}} \sum y_j^{2}$ and $\langle \theta^{2} \rangle = \frac{1}{N_{\text{ele}}}\sum \theta_j^{2}$ characterizing the spatial and angular distributions respectively, while $\langle y \theta \rangle = \frac{1}{N_{\text{ele}}} \sum y_j\theta_j$ represents their covariance; this formulation effectively isolates the fundamental bunch quality metric by subtracting the geometric artifacts induced by bunch tilt from the product of spatial and angular variances, where $N_{\text{ele}}$ is the total number of electrons.The simulated data yielded a spatial emittance of $0.49 \times 10^{-3}$\,mm$\cdot$rad for the electron bunch at $t=27\mathrm{T_0}$.
		
		To investigate the temporal evolution of the electron bunch emittance, we assigned unique identifiers to the macro-electrons in this region (where each macro-electron represents $6.6 \times 10^5$ physical electrons) for tracking their temporal variations. This approach provided complete data on both the spatial and momentum evolution of these electrons. Figure.~\ref{fig4}(a). shows the temporal evolution of the electron bunch emittance. It can be observed that although the emittance increases with time, the maximum emittance remains at a relatively low level, indicating excellent bunch collimation over multiple cycles.
		
		Additionally, we present the spatial distributions of electron positions at different time instances through scatter plots as shown in Figure.~\ref{fig4}(b). It displays the electron distributions at three specific time points: $t=27\mathrm{T_0},t = 30\mathrm{T_0},  t = 32\mathrm{T_0}$, revealing a significant spatial diffusion of electrons over time.   At the early stage of the interaction ($t=27\mathrm{T_0}$), an electron-ion double layer is formed by the ponderomotive force, and the laser radiation pressure is balanced by the pressure of the electrostatic field. This prevents electrons from Coulomb explosion and ponderomotive expansion, resulting in a concentrated electron distribution in space. However, at later stages of the interaction ($t=30\mathrm{T_0}$ and $t=32\mathrm{T_0}$), as the intensity of the harmonic pulse increases, the thin target configuration employed does not fully satisfy the conditions required for complete double layer formation. As a result, electrons are expelled from the ions by the harmonic pulse and undergo direct acceleration through radiation pressure. These electrons are ponderomotively scattered in the perpendicular direction and experience Coulomb explosion, ultimately leading to a more dispersed electron distribution in space, as shown in Figure.~\ref{fig4}(b). Furthermore, the figure also demonstrates that the majority of electrons remain within the transverse focal spot region and the leading portion of the electric field at $t=30\mathrm{T_0}$ and $t=32\mathrm{T_0}$. These electrons exhibit continuous energy gain, indicating sustained acceleration by the high-order harmonic electric field through direct acceleration.

		As shown in Figures.~\ref{fig4}(c) and (d), we present the distributions of the Lorentz factor ($\gamma$) for electrons at $t=27\mathrm{T_0}$ and $t=30\mathrm{T_0}$, respectively. At $t=27\mathrm{T_0}$, the distribution exhibits a relatively narrow peak, indicating a more uniform energy state among the electrons at this early stage of interaction. In contrast, by $t=30\mathrm{T_0}$, the Lorentz factor distribution broadens significantly, with a clear extension toward higher $\gamma$ values. The results demonstrate that the higher $\gamma$ values are predominantly concentrated within a specific range, which is fully consistent with our previous analysis. This concentration suggests that a distinct acceleration mechanism dominates during this phase, likely linked to the interplay between the harmonic electric field and the spatial distribution of the electron bunch. The consistency between these observations and earlier findings reinforces the reliability of the simulation model and supports the hypothesis that the acceleration process is both coherent and reproducible under the given experimental conditions.

		We plotted the angular energy distribution of electrons at these three time points. Figures.~\ref{fig5}(a), (b), and (c). correspond to the angular energy distributions of electrons at $t=27\mathrm{T_0}$, $t=30\mathrm{T_0}$, and $t=32\mathrm{T_0}$, respectively. The $180^\circ$ direction denotes the laser propagation direction ($x$-axis), and the radial direction represents the electron energy. At time $t=27\mathrm{T_0}$, the accelerated electron bunch is primarily forward-propagating after interacting with the reflected laser (harmonic pulse), and its maximum energy exceeds 35 MeV, with a relatively small opening (divergence) angle of $10^\circ$, as shown in Figure.~\ref{fig5}(a). However, at times $t=30\mathrm{T_0}$ and $t=32\mathrm{T_0}$, the electron bunch energy increases to over 50 MeV, while the opening angle also increases, as shown in Figures.~\ref{fig5}(b) and (c). This behavior is consistent with the spatial broadening of the electron density distribution in space [see Figures.~\ref{fig4}(b)--(d)].
		
		It can also be observed from Figures.~\ref{fig5}(a), (b) and (c). that electrons gradually deviate from the $x$-axis (forward direction) over time. At the beginning ($t=27\mathrm{T_0}$), the deviation angle is only approximately $\sim 6^\circ$, with higher-energy electrons being more concentrated near the optical axis, as shown in Figure.~\ref{fig5}(a). The primary reason is that during this period, the amplitude of the electrostatic force remains sufficiently strong to balance the ponderomotive force, thereby enabling effective electron confinement along the propagation direction. However, at later time points, the combined effects of the self-generated magnetic field induced by electron perturbations cause most electrons with velocity in the $x$-direction to experience a magnetic force in the $y$-direction, leading to a deviation from the optical axis. Over time, this cumulative effect results in transverse divergence of the electron bunch. By $t=32\mathrm{T_0}$, as shown in Figure.~\ref{fig5}(c), the angular energy distribution of electrons exhibits a significant deviation from the $x$-axis, reaching approximately $\sim 30^\circ$.

\section{Discussion}\label{SecIV}
	\subsection{Direct interantion of a Gaussian laser}\label{SecIV--cause}
		In order to compare the physical mechanisms of attosecond electron bunch generation using a Gaussian laser pulse directly versus a high-order harmonic pulse, we conducted an additional simulation in which a Gaussian laser not reflected by the plasma mirror is directly applied to the thin target. The simulation time and spatial domain are identical to those used in previous simulations. In this additional simulation, the laser front reaches the target at $t=15\mathrm{T_0}$, and after two laser cycles, the main portion of the laser interacts with the thin target. Consequently, an electron bunch is generated behind the thin target, following a mechanism similar to that observed in the earlier case involving the high-order harmonic pulse.
		The origin of the ultrashort electron bunch can be attributed exclusively to the influence of the laser electric field. As observed from Figures.~\ref{fig2}(a) and (b), a conventional Gaussian laser pulse exhibits a temporally symmetric, multi-cycle sinusoidal electric field oscillation. During its interaction with the thin target, the electric field continuously injects electrons into the accelerating region across multiple optical cycles. Due to the symmetry of the field, electrons accelerated during the first half-cycle may be decelerated during the subsequent half-cycle. Electrons extracted by the electric field at different phases experience varying acceleration conditions, ultimately forming an elongated electron bunch in space, characterized by an extended temporal duration lower charge.
		
		For a high-order harmonic pulse, the temporal electric field structure undergoes a fundamental transformation, exhibiting pronounced sub-cycle modulation and several isolated high-amplitude peaks. These peaks are characterized by an extremely narrow full-width at half maximum (FWHM). More importantly, the electric field becomes asymmetric, featuring an extended negative half-cycle, as illustrated by the red curve in Figure.~\ref{fig2}(a). This asymmetry indicates that the harmonic pulse energy is delivered within an extremely short time interval, enabling the efficient pulling and acceleration of electrons to higher energies within a reduced temporal window. Consequently, the resulting electron bunch is also extremely short, with its duration approaching the width of the driving field peak.

		Furthermore, the spectrum reveals that the high-harmonic signal [red curve in Figure.~\ref{fig2}(b)] is significantly broadened into the high-frequency region and remains continuous across the entire bandwidth. According to the time-bandwidth product ($\Delta\nu\Delta t \gg 1$), the pulse width is inversely proportional to the spectral bandwidth. In contrast, the narrow spectral width of a conventional Gaussian laser [blue curve in Figure.~\ref{fig2}(b)] fundamentally limits the minimum pulse duration it can support, typically restricting it to the femtosecond regime due to its Fourier transform limit. The broad bandwidth of high harmonics, however, enables the synthesis of attosecond pulses, making high-harmonic generation an effective method for producing such ultrashort pulses, similar to extracavity spectral amplification techniques. Therefore, the high harmonic itself, as a driving source, possesses the intrinsic physical capability to generate ultra-short electron bunchs. This characteristic directly constrains the electron acceleration process to occur within a time window on the order of a single attosecond, ultimately leading to the formation of ultra-short electron bunches.

	\subsection{Effects of different driving laser intensities}
\begin{figure}[ht]\suppressfloats
			\includegraphics[scale=0.60]{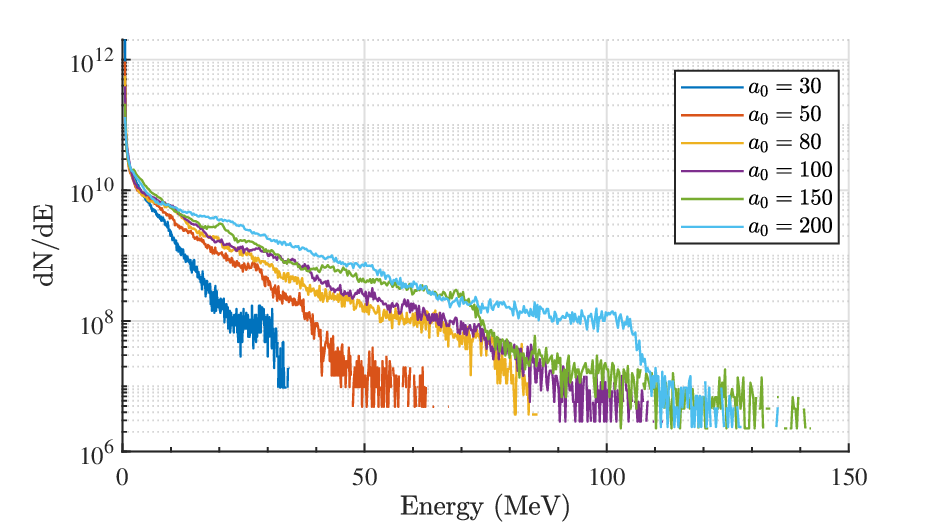}
			\caption{Electron bunch energy spectra at $t=27\mathrm{T_0}$ under different laser intensities: $a_0 = 30, 50, 80, 100, 150,$ and $200$.}
			\label{fig6}
		\end{figure}

		To investigate the effects of different laser intensities on the characteristics of ultrashort electron bunchs, we conducted simulations using lasers with normalized intensities of $a_0 = 30, 50, 80, 100, 150, 200$. The electron bunch spectra under different laser intensities at $t=27\mathrm{T_0}$ are presented in Figure.~\ref{fig6}.
		\begin{figure}[ht]\suppressfloats
			\includegraphics[scale=0.60]{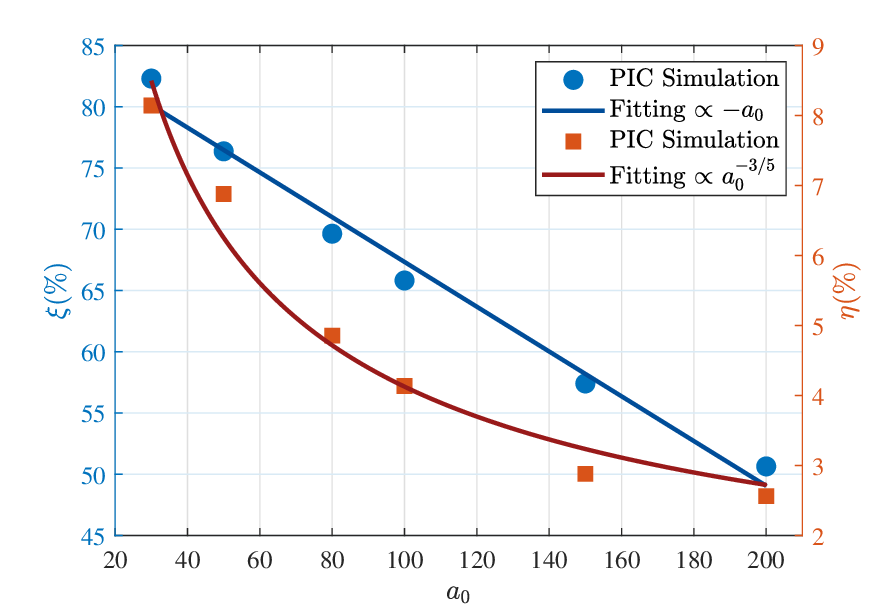}
			\caption{The energy conversion efficiencies from the laser to the harmonic pulse and from the harmonic pulse to the electron bunch. The red solid circles and blue squares represent the data obtained from PIC simulations, and the solid lines correspond to the respective fitting curves.}
			\label{efficency}
		\end{figure}
\begin{figure*}[ht]\suppressfloats
	\includegraphics[scale=0.25]{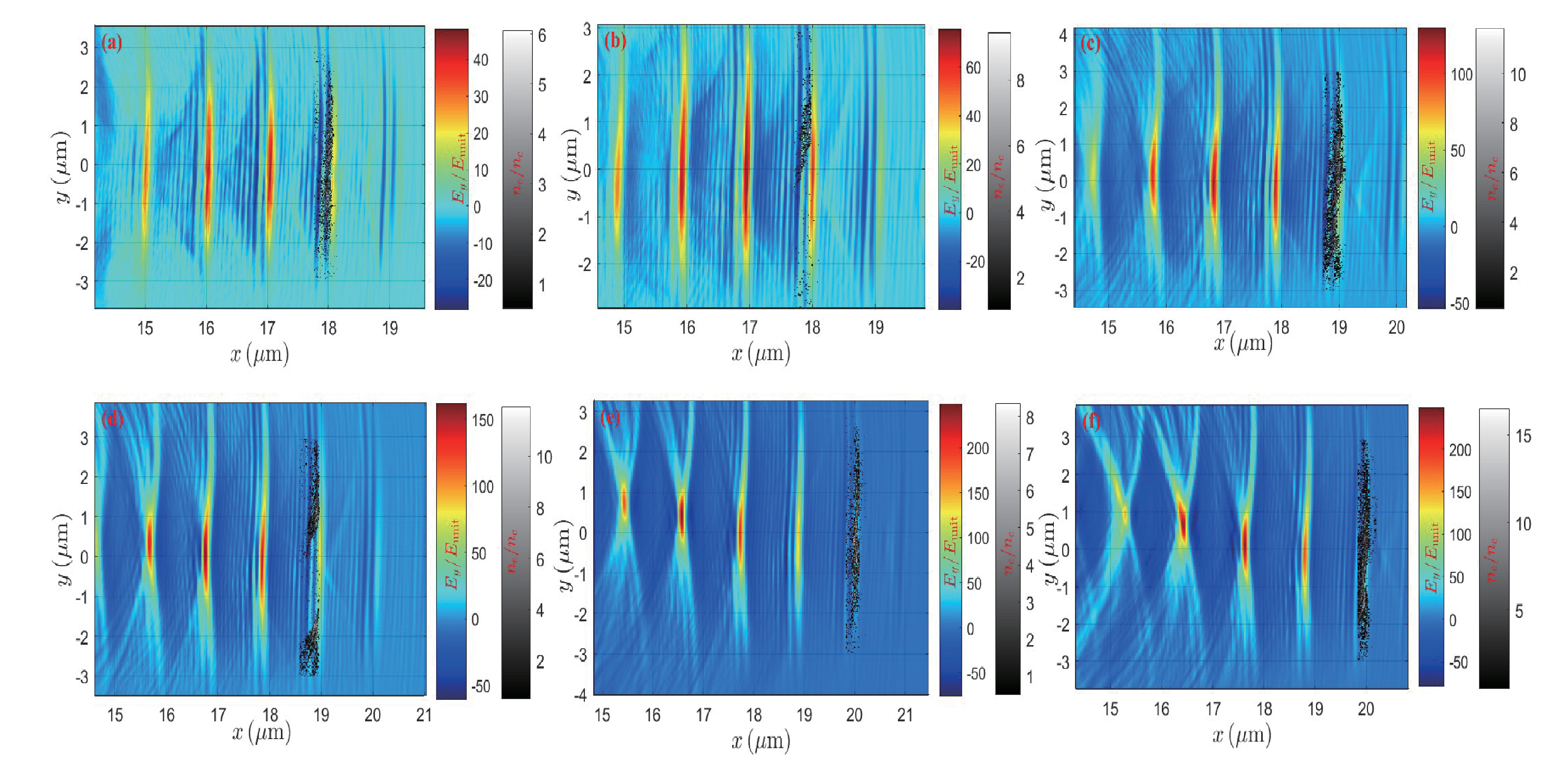}
	\caption{Distributions of the electric field and electron bunch density at $t=27\mathrm{T_0}$. Panels (a)--(f) correspond to driving laser intensities of $a_0 = 30$, $50$, $80$, $100$, $150$, and $200$, respectively.}
	\label{fig8}
\end{figure*}
As can be observed, the electron energy gradually increases with the laser intensity. However, when the laser intensity reaches $a_0 = 200$, the maximum electron energy does not exhibit a significant increase compared to the case at $a_0 = 150$. This is primarily due to the fact that, in our simulation setup, the number of electrons within the focal spot radius remains constant. As the laser intensity increases, a larger fraction of electrons in this region is accelerated. This behavior is evident from the figure, which shows that at an energy level of $20~\mathrm{MeV}$, the number of electrons increases with increasing intensity. We also conducted simulations for a laser intensity of $a_0 = 300$, and found that although the electron cut-off energy can reach approximately $250~\mathrm{MeV}$, the harmonic characteristics of the reflected laser are significantly degraded. At such high intensities, the laser may easily induce pre-plasma breakdown, thereby adversely affecting the higher harmonic spectrum and finall electron bunch characteristics.

\begin{table*}[t]
    \centering
    \caption{Electron bunch parameters for driving laser intensities}
    \label{tab:laser_data}
    	\begin{tabular}{c l S[table-format=1.2] S[table-format=1.1e-2] S[table-format=3.0]}
    		\mytoprule
    		\textbf{Laser Intensity} &
    		\textbf{bunch Position (\si{\micro\meter})} &
    		\textbf{Charge (\si{nC})} &
    		\textbf{Emittance (\si{mm \cdot mrad})} &
    		\textbf{Duration (\si{as})} \\
    		\mymidrule
    		30  & 17.8--18.2 & 0.38 & 4.9e-4 & 100 \\
    		50  & 17.7--18.1 & 0.37 & 8.5e-4 & 100 \\
    		80  & 18.7--19.1 & 0.99 & 4.5e-4 & 200 \\
    		100 & 18.6--19.0 & 0.68 & 7.0e-4 & 233 \\
    		150 & 19.8--20.2 & 0.59 & 2.5e-4 & 266 \\
    		200 & 19.8--20.2 & 1.15 & 2.7e-4 & 400 \\
    		\mybottomrule
    	\end{tabular}
\end{table*}

Furthermore, we calculated the energy conversion efficiencies at various laser intensities, as shown in Figure.~\ref{efficency}. Here, $\xi$ denotes the energy conversion efficiency from the driving laser to the high-order harmonics after complete reflection, while $\eta$ represents the efficiency of energy transfer from the high-order harmonics to the electron bunch. The results indicate that $\xi$ exhibits a linear decay trend, whereas $\eta$ follows $a_{0}^{-3/5}$ decay. The energies used for this analysis were recorded after the completion of high-harmonic pulse generation and at the electron bunch time of $t=27\mathrm{T_0}$. As shown in the figure, the overall energy conversion efficiency gradually decreases with increasing laser intensity. At $a_0 = 200$, nearly half of the energy is lost, indicating a substantial reduction in energy utilization efficiency.

Table~\ref{tab:laser_data} summarizes the position, charge quantity, emittance and duration  of the electron bunch at $t=27\mathrm{T_0}$ under different laser intensities. It can be observed that the electron emittance remains relatively stable across all laser intensity levels. However, the electron charge quantities exhibit notable variation, with the maximum charge reaching approximately 1 nC at $a_0 = 200$, although this is accompanied by a relatively low laser energy conversion efficiency at this intensity.
Regarding the variations in charge quantities under different intensities: at lower intensities, electrons are primarily trapped in the negative electric field region of the harmonic pulse and accelerated in the forward direction. Under these conditions, the electron bunch remains confined within the electromagnetic field of the harmonic pulse, as indicated in Table~\ref{tab:laser_data} (bunch position) and Figure.~\ref{fig8}(a). This leads to the formation of a compact electron bunch, as illustrated in Figures.~\ref{fig3}(b) and \ref{fig8}(a).
When the intensity increases to $a_0 = 50$, the reflected harmonic pulse becomes sufficiently strong to disrupt the thin target. The transverse ponderomotive force induced by the laser results in bunch splitting, leading to the formation of two distinct electron bunchs, as shown in Figure.~\ref{fig8}(b). This phenomenon contributes to a decrease in the overall electron bunch charge, as observed in Table~\ref{tab:laser_data}.

At a laser intensity of $a_0 = 80$, the harmonic pulse is sufficiently intense to enable electrons within the laser spot to undergo direct light sail acceleration, resulting in a high-charge electron bunch with a charge of 0.99 nC, as depicted in Figure.~\ref{fig8}(c).
As the driving laser intensity further increases to $a_0 = 100$ and beyond, the target used for harmonic generation undergoes deformation and no longer satisfies the optimal conditions for efficient harmonic generation. Consequently, the reflected pulse deviates from the optical axis, as illustrated in Figures.~\ref{fig8}(c)–(d). This deviation leads to irregular variations in both the bunch charge and duration, as presented in Table~\ref{tab:laser_data}.

From the perspective of high-harmonic generation mechanisms, it is well established that the properties of the electron layer acting as the reflecting target—such as the density gradient—have a significant impact on the emission characteristics of high-order harmonics. For the pre-plasma parameters employed in our model, a laser with an intensity corresponding to $a_0 = 30$ achieves a higher degree of coupling efficiency between the laser and the target compared to lasers with higher intensities.
		
	\subsection{Effetes of different thin target positions}
		We also investigated the influence of thin targets placed at different positions on the electron bunch characteristics. In the previous simulations, our thin target was positioned at $\SI{15}{\micro\meter}$, which coincides with the location of complete laser reflection. This placement ensured its interaction with a fully formed high-harmonic field in a quasi-far-field propagation state. To further explore the potential influence of the early near-field characteristics of high-harmonic generation (HHG) on electron dynamics, we systematically performed a series of parameter scan simulations, relocating the target to positions at $L = \SIlist{10;11;12;13;14}{\micro\meter}$. These regions are the peaks of the electric field corresponding to each other in space after the harmonic reflection is complete.

		From the two-dimensional electron number density distribution, it can be observed that although the high-order harmonics are not fully formed at this stage, the laser reflection from the pre-plasma has already broadened the frequency spectrum, resulting in a wide-band spectral component that spatially forms an electron bunch-like structure.
		Table~\ref{tab:target_results}  shows the charge, emissivity and duration of the electron beam at different positions of the thin target. It can be seen that the duration of the electron beam is not much different as the thin target position gradually moves back, but the amount of charge gradually increases. When the thin target is positioned too far forward, the laser-plasma modulation process remains incomplete, resulting in highly unstable electric fields within this region. The electric field simultaneously interacts with both the reflecting target and the thin target, and the coupling between these interactions prevents efficient acceleration of the electron bunch.  The further back the thin target position, the higher harmonics are about to form, and from the analysis of Figure.~\ref{fig2}(a), the electric field at $\SI{13}{\micro\meter}$ after reflection has stabilized.
		\begin{table}[t]
    			\centering
    			\caption{Electron bunch parameters for different target positions}
   				 \label{tab:target_results}
    			\setlength{\tabcolsep}{3pt}
    			\renewcommand{\arraystretch}{1.0}
    			\begin{tabular}{@{}c S[table-format=1.2] S[table-format=1.1e-2] S[table-format=2.2]@{}}
    			\hline
    			\textbf{Position} & \textbf{Charge} & \textbf{Emittance} & \textbf{Duration} \\
   				 \textbf{(\si{\micro m})} & \textbf{(\si{nC})} & \textbf{(\si{mm \cdot mrad})} & \textbf{(as)} \\
    			\hline
    			10 & 0.27 & 5.5e-4 & 133 \\
    			11 & 0.21 & 4.2e-4 & 166 \\
    			12 & 0.26 & 2.6e-4 & 266 \\
    			13 & 0.31 & 3.9e-4 & 200 \\
    			14 & 0.34 & 4.1e-4 & 166 \\
				15 & 0.38 & 4.9e-4 & 100 \\
    			\hline
    			\end{tabular}
		\end{table}

\section{CONCLUSION}\label{SecV}
		The high-order harmonics generated by extracavity spectrum amplification are one of the current methods to obtain attosecond pulses,
		we employed a Gaussian laser interacting with a pre-formed plasma target to generate these harmonics, which served as the driving source. This approach enabled the production of electron bunches with significantly shorter durations compared to those obtained through the direct interaction of femtosecond lasers with solid-density thin targets. The achieved bunch duration reached 100 as, marking a substantial reduction in the pulse width of laser-solid-target-generated electron bunch. Furthermore, the generated bunches exhibited favorable quality in terms of emittance and energy-angular distribution, while maintaining a substantial charge on the order of nanocoulombs.
		
		At the same time, we also studied the effect of laser intensity on the electron beam, because the high harmonics generated by the laser and the pre-plasma in the reflection stage are strongly dependent on the influence of laser and pre-plasma target parameters, in order to improve the coupling efficiency of the two, we found that this process is not suitable for lasers with too high intensity. We also study the location of thin targets, in order to obtain an electron beam with a large amount of charge, the harmonics must be fully reflected. These simulations provide an important theoretical basis and optimization direction for the generation of electron bunch with high collimation and large charge. By systematically adjusting the thin target position and laser parameters, we further optimize the key performance indicators of the electron beam, laying a solid foundation for realizing high-quality electron beam sources based on laser plasma interaction, which has important application value in the fields of ultrafast electron diffraction, new accelerator technology, and attosecond science.

\begin{acknowledgments}
This work was supported by the National Natural Science Foundation of China
(NSFC) (Grant Nos. 12265024, 12564046, and 12375240), and the Special Training Program of Science and Technology Department of Xinjiang China (Grant No. 2024D03007). Yang He also appreciate the Outstanding Graduate Innovation Project of Xinjiang University with No. XJDX2025YJS153.
\end{acknowledgments}

\end{document}